\documentclass[11pt]{article}
\pdfoutput=1
\usepackage{jcappub,amsmath,amssymb}
\usepackage{graphics}
\usepackage{enumerate}
\usepackage{todonotes}
\usepackage{titlesec}
\titleformat{\paragraph}[runin]{\normalfont\normalsize\bfseries}{\theparagraph}{0pt}{\hspace{0.85cm}}[]

\usepackage[retainorgcmds]{IEEEtrantools}

\definecolor{hyperref}{RGB}{026,028,087}

\newcommand{\be}{\begin{equation}}
\newcommand{\ee}{\end{equation}}

\newcommand{\uv}[1]{\hat{\mathbf{#1}}}

\graphicspath{{./}}
\notoc

\begin{document}

\title{An anisotropic universe due to dimension-changing vacuum decay}

\author[a]{James H. C. Scargill}

\affiliation[a]{Theoretical Physics, University of Oxford, DWB, Keble Road, Oxford, OX1 3NP, UK} 

\emailAdd{james.scargill@physics.ox.ac.uk}

\abstract{
In this paper we consider the question of observational signatures of a false vacuum decay event in the early universe followed by a period of inflation; in particular, motivated by the string landscape, we consider decays in which the parent vacuum has a {\it smaller} number of large dimensions than the current vacuum, which leads to an anisotropic universe. We go beyond previous studies, and examine the effects on the CMB temperature and polarisation power spectra, due to both scalar and tensor modes, and consider not only late-time effects but also the full cosmological perturbation theory at early times. We find that whilst the scalar mode behaves as one would expect, and the effects of anisotropy at early times are sub-dominant to the late-time effects already studied, for the tensor modes in fact the the early-time effects grow with multipole and can become much larger than one would expect, even dominating over the late-time effects. Thus these effects should be included if one is looking for such a signal in the tensor modes.
}


\maketitle
\newpage

\section{Introduction} \label{sec-intro}

The study of false vacuum decay in quantum field theory is a topic with a long history, dating back to the initial work of Callan, Coleman, and de-Luccia \cite{Coleman:1977py, Callan:1977pt, Coleman:1980aw}; the basic idea is that if the theory has multiple vacuum states, and initially is in one with higher energy, henceforth the false vacuum, then it is possible to tunnel to a state in which there is a `bubble' of lower-energy (henceforth, true) vacuum sitting in the remaining region of false vacuum, the two regions being separated by a transitionary bubble wall. This bubble then proceeds to grow in size until, in the asymptotically far future, the whole universe is in the true vacuum.

Within the bubble it is possible to construct the coordinate system of an open FLRW universe (with the bubble wall at infinity) \cite{Coleman:1980aw}, which leads to the exciting possibility that if our universe were open, it could have arisen out of such a false vacuum decay event, followed by a short period of inflation (in part to dilute away the spatial curvature, which would dominate initially). Much work was done concerning this idea in the early `90's, for example see \cite{Bucher:1994gb, Yamamoto:1996qq, Garriga:1998he, Linde:1999wv}, 
and there has since been a resurgence of interest due the idea of the multiverse, for example \cite{Freivogel:2005vv, MarchRussell:2006mj, Yamauchi:2011qq, Bousso:2013uia, White:2014aua}, since if such a landscape of possible metastable vacuum states exists (e.g. \cite{Bousso:2000xa, Feng:2000if, Susskind:2003kw}), then tunnelling events between these states would seem inevitable. Thus a tunnelling event in the origin of our universe is, in some sense, a prediction of the landscape hypothesis, and it behoves us to investigate the observational consequences. 

In particular, this paper examines the consequences of tunnelling events in which the number of large dimensions in the false vacuum is {\it smaller} than the number of large dimensions in the true vacuum, as could occur for example if the tunnelling field is the modulus of one (or more) of the dimensions. This possibility is motivated in part by string theory; the string landscape arises out of the myriad ways to compactify any of the ten dimensions required for the consistency of the theory (including the possible flux quanta that can exist on closed sub-manifolds, {\it i.e.}, homology `cycles' of various dimensionality, of the compactification), in order to generate a low energy effective theory in a smaller number of dimensions.  Including the many discrete choices of fluxes there are generically vastly more ways to compactify a larger number of dimensions than a smaller number.  This leads one to expect that there are more vacua in the landscape which have a smaller number of large, apparently non-compact, spatial dimensions. (See \cite{SchwartzPerlov:2010ne, Chung:2012fa} for attempts to tackle the measure problem in such a scenario.) Thus in the the (string) landscape a tunnelling event which increases the number of large dimensions is a motivated and intriguing possibility \cite{Linde:1988yp, Giddings:2004vr, Carroll:2009dn, BlancoPillado:2009di, BlancoPillado:2009mi}. This is not to say that such events are exclusive to string theory, for example even the Standard Model may support lower dimensional vacua \cite{ArkaniHamed:2007gg, Arnold:2010qz}. 

Finally, many other approaches to quantum gravity besides string theory suggest that space-time may be of lower dimensionality at high energies \cite{Carlip:2009km, Calcagni:2009kc}. This could manifest itself observationally in many ways \cite{Anchordoqui:2010er, Mureika:2011bv}, and in particular in a scenario similar to that considered here, with inflation after a dimensional transition \cite{Rinaldi:2010yp}.

Such dimension-changing false vacuum decay events and their observational signatures have been considered before. The key point is the that the universe that results will have large background anisotropy immediately after its creation (at least in the $2+1 \to 3+1$ and $1+1 \to 3+1$ cases).  The background anisotropy reasserts itself during the matter dominated epoch, and the effect of this on the CMB temperature power spectrum was examined in \cite{Graham:2010hh}, while in \cite{BlancoPillado:2010uw, Adamek:2010sg}, and more recently \cite{Blanco-Pillado:2015dfa}, the tunnelling process itself was investigated, along with the effect on the scalar power spectrum in the limit of no metric fluctuations. In this paper we extend this to the polarisation power spectra, specifically going beyond previous work on the effects of early-time anisotropy by performing full cosmological perturbation theory allowing us to examine the effect on the tensor perturbation modes. 

The paper is structured as follows: section \ref{sec-model} discusses in more detail the model which we are considering, as well as the background evolution of the universe; section \ref{sec-early} considers perturbation theory in a universe with background anisotropy, derives the primordial power spectra for the case of $2+1 \to 3+1$, and analyses how to obtain CMB power spectra, ignoring the effect of late-time anisotropy, which is then considered in section \ref{sec-late}. Section \ref{sec-results} then gathers together the results, compares the effects, and comments on observability. Finally section \ref{sec-conc} concludes. Four appendices contain more detail on the various calculations presented in the main body of the paper.

\section{An anisotropic universe} \label{sec-model}

We here focus on the case of $2+1 \to 3+1$ and $1+1 \to 3+1$ tunnelling events resulting in our present universe but with
anisotropy of a particular form. 
The universe inside the bubble (our universe) is spatially homogenous, but anisotropic, with the possibility of curvature in two of the spatial dimensions and cosmic shear between those and the third spatial dimension; its metric is
\be
ds^2 = -dt^2 + a(t)^2 \left( \frac{dr^2}{1- \kappa r^2} + r^2 d\phi^2 \right) + b(t)^2 dz^2. \label{background metric}
\ee
The sign of $\kappa$ depends on the origin of the $(r,\phi)$ dimensions. If they are `old', i.e. the $z$ dimension is the one which has recently decompactified (so the transition was $2+1 \to 3+1$), then $\kappa = -1$, since the coordinate system inside the bubble (with the `new' $z$ dimension suppressed) must be open.  In this case \eqref{background metric} describes a Bianchi III universe. On the other hand if $(r,\phi)$ are the new dimensions, and the transition was $1+1 \to 3+1$, then $\kappa \in \{ -1, 0, 1 \}$ depending on the details of their previous compactification (e.g. compactification on $S^2$ yields $\kappa = 1$).  In this case \eqref{background metric} describes a Bianchi III, Bianchi I, or Kantowski-Sachs universe respectively.

From the $tt$ and $zz$ components of the Einstein equations one finds
\begin{align}
H_a^2 + 2 H_a H_b + \frac{\kappa}{a^2} &= \frac{1}{M_\text{Pl}^2} \rho, \\
2 \dot{H}_a + 3 H_a^2 + \frac{\kappa}{a^2} &= - \frac{1}{M_\text{Pl}^2} p_z,
\end{align}
which are then supplemented with the equation of motion for matter, or the $rr$ or $\phi\phi$ component of the Einstein equations.  Here $\rho$ and $p_z$ are the $tt$ and $zz$ components respectively of the stress tensor $T^\mu_{\phantom{1}\nu}$, $M_\text{Pl}$ is the reduced Planck mass, and $H_a^2 \equiv (\dot{a}/a)^2$, $H_b^2 \equiv (\dot{b}/b)^2$.
We shall assume that inflation is driven by a single scalar field $\phi$, with potential $V$, in which case during inflation the equations become
\begin{align}
H_a^2 + 2 H_a H_b + \frac{\kappa}{a^2} &= \frac{1}{M_\text{Pl}^2} \left( \frac{1}{2} \dot{\phi}^2 + V \right), \\
2 \dot{H}_a + 3 H_a^2 + \frac{\kappa}{a^2} &= \frac{1}{M_\text{Pl}^2} \left( - \frac{1}{2} \dot{\phi}^2 + V \right), \\
\ddot{\phi} + \left( 2H_a + H_b \right) \dot{\phi} + V' &= 0.
\end{align}
The initial conditions depend on the origin of the dimensions, with the scale factor of the `old' dimensions starting from zero, and we assume that the scale factor for the new dimensions starts at rest from some finite value (which would be expected if the tunnelling field were the modulus for the extra dimension) along with the inflaton.

The evolution of the background has been studied in \cite{Graham:2010hh} and will be briefly recounted here. Given the boundary conditions mentioned above, immediately after nucleation of the bubble the behaviour depends on which dimensions are new: in the $2+1 \to 3+1$ case we have $a(t) \approx t + \frac{V}{18 M_\text{Pl}^2}t^3 + \mathcal{O}
(t^4)$ and $b(t) \approx b(0)\left(1 + \frac{V}{6 M_\text{Pl}^2}t^2 + \mathcal{O}
(t^4) \right)$; the $1+1 \to 3+1$ case depends on the sign of the curvature. As inflation proceeds the anisotropy in expansion rate will be driven to zero, as will the anisotropic curvature parameter
\be
\Omega_k = \frac{\kappa}{a^2 H_a^2}.
\ee
In the radiation and matter dominated epochs, like any curvature, this will begin to grow, and will source a corresponding anisotropy in the expansion rate (in \cite{Graham:2010hh} it was shown that $\frac{H_a - H_b}{H_a} \propto - \Omega_k$).

It is important to note that there are {\it two} types of anisotropy in \eqref{background metric}: shear anisotropy due to the fact that $H_a \neq H_b$, and curvature anisotropy due to the fact that only two of the three spatial dimensions are curved. 
Thus the observational consequences of this model can arise both from anisotropy at early times, which we examine in section \ref{sec-early}, and from anisotropy at late times, which we discuss in section \ref{sec-late}.

\section{Early-time anisotropy} \label{sec-early}

The effect of anisotropy at early times is to modify the behaviour of the perturbations to the inflaton and metric, whose fluctuations will eventually be imprinted upon the CMB. Thus we need to extend the usual cosmological perturbation theory to deal with universes of the form \eqref{background metric}; in order to do this we will closely follow \cite{Gumrukcuoglu:2007bx} in which this was done for universes with anisotropic shear, but no curvature. (See also \cite{Pereira:2007yy} for the same, and \cite{Zlosnik:2011iu} in which is examined the case of anisotropic curvature, but no shear.)

We immediately note two things:
\begin{enumerate}[(i)]
\item Since we have two scale factors, it is not obvious how we should define conformal time, and in fact the appropriate definition will depend on the specific case (i.e. which are the new dimensions) under consideration; thus for now we just define $c(t) d\eta = dt$, where $c(t)$ is some function of $a$ and $b$.
\item As we no longer have a residual $SO(3)$ symmetry on the spatial slices, we cannot use the action of this to classify the perturbations, and instead must use the $SO(2)$ symmetry of $(r, \phi)$; thus the perturbations will be $SO(2)$ scalars (1 degree of freedom) and vectors (also 1 d.o.f. since it has two components and is required to be transverse).\footnote{Note that $SO(2)$ tensors have four components and are required to be symmetric, traceless, and transverse, which eliminates all the d.o.f. and thus they do not appear.}
\end{enumerate}
The gory details of anisotropic perturbation theory are relegated to appendix \ref{app-perturbation theory}, and here we simply state the results. Of the $SO(2)$ vectors only one survives, and comparison with the isotropic case\footnote{Note that due to the presence of anisotropic curvature, the isotropic limit is not just $a = b$, but also requires sending $\Omega_k \to 0$.} reveals this to be the analogue of the tensor perturbation $h_\times$, and so we denote it likewise. The quadratic action for it is given by
\be
\mathcal{S}^{(2)}_\times = \frac{1}{2} \int d\eta\ d^2q\ dk_3 \sqrt{\gamma} \left( | h_\times ' |^2 - \omega_\times^2 | h_\times |^2 \right). \label{S2x}
\ee 
Here $-q^2$ is the eigenvalue of the 2-dimensional Laplacian\footnote{Note that it satisfies the following relations depending on the value of $\kappa$ \cite{Vilenkin_Smorodinskii}:
\be
q^2 = 
\begin{cases}
k_2^2 + \frac{1}{4} & \text{if } \kappa = -1, \\
k_2^2 & \text{if } \kappa = 0, \\
l(l+1) & \text{if } \kappa = 1,
\end{cases}
\ee
where $k_2 \in \mathbb{R}$ and $l \in \mathbb{N}^0$. Though for compactness we will also write $q^2 = k_2^2$ when $\kappa = 1$.\label{k_2 definition}} on $(r,\phi)$, $-k_3^2$ is the eigenvalue of the 1-dimensional Laplacian for the $z$ dimension, $\gamma$ is the determinant of the conformal $(r,\phi)$ metric, and the effective, conformal-time-dependent mass is given by
\be
\omega_\times^2 = c^2 \left( \frac{q^2}{a^2} + \frac{k_3^2}{b^2} \right) - 2 \kappa \left( \frac{c}{a} \right)^2 - \frac{\left[ \sqrt{\frac{a^2}{b c} \left( \frac{q^2}{a^2} + \frac{k_3^2}{b^2} \right)^{-1} } \right]''}{\sqrt{\frac{a^2}{b c} \left( \frac{q^2}{a^2} + \frac{k_3^2}{b^2} \right)^{-1} }}. \label{wx}
\ee
Note that in the isotropic limit this reproduces the expected result, $k^2 - \frac{a''}{a}$.

Two $SO(2)$ scalar perturbations survive, which are the analogues of the isotropic scalar mode $v$ (the Mukhanov-Sasaki variable) and the tensor mode $h_+$. These two are coupled and their action takes the form
\be
\mathcal{S}^{(2)}_{v, +} = \frac{1}{2} \int d\eta\ d^2q\ dk_3 \sqrt{\gamma} \left( H'^\dagger K H' + H'^\dagger \tilde{K} H - H^\dagger M H \right), \label{S2v+}
\ee 
where $H^\mathrm{T} = ( \varphi , \Sigma )$, and the matrices $K, \tilde{K}$, and $M$ are given in Appendix \ref{app-perturbation theory}. In order to find the precise relationship between $\{\varphi, \Sigma\}$, and $\{v, h_+\}$, the mixing in the first two terms must be removed through a suitable field redefinition. Whilst this can be done in principle, and it is possible to do it in closed form in the absence of curvature \cite{Gumrukcuoglu:2007bx}, the correct procedure for the general case eludes us here. Fortunately, in the simplified case we consider below the demixing will happen automatically.

We can now also answer the question of how to define conformal time. The requirement is that, for each mode, the adiabatic vacuum state (the generalisation of usual vacuum state in Minkowski space),
\be
u(\eta) = \frac{1}{\sqrt{2 \omega}} e^{-i \int^\eta \omega(\eta') d\eta'}, \label{adiabatic vacuum}
\ee
is a solution of the equations of motion as $t \to 0$. This corresponds to $\omega' \ll \omega^2$, and $\omega^2 \geq 0$ as $t \to 0$. By 
inspecting \eqref{wx} we see that this is satisfied when $c$ is the scale factor of the old dimensions (i.e. those for which the scale factor goes to zero as $t$ does), and the same is true for the modes in \eqref{S2v+}.

\subsection{Primordial power spectra}

We are interested in the power spectrum of the fluctuations at the end of inflation.  Since inflation efficiently isotropises the background, the modes become their isotropic counterparts; therefore in the remainder of the paper we will refer to them as such. The relation of the primordial power spectra to the expectation value of the modes is the same as in the isotropic case, i.e. for the comoving curvature perturbation $\mathcal{R} = \frac{H}{a \dot{\phi}} v$,
\be
\langle \mathcal{R}(\mathbf{k}) \mathcal{R}^*(\mathbf{k}')  \rangle = \frac{(2 \pi)^3}{k^3} \delta(\mathbf{k} - \mathbf{k}') P_\mathcal{R}(k,\mu), \label{power spec def}
\ee
and similarly for the tensor perturbations $\psi_{+,\times} = \frac{1}{a} h_{+,\times}$. Here $k^2 = k_2^2 + k_3^2$ (see footnote \ref{k_2 definition} for the definition of $k_2$) and $\mu  = \uv{k} \cdot \uv{z}$ is the cosine of the angle that $\mathbf{k}$ makes with the $z$ direction.
The fact that the power spectrum only has this dependence on the direction of the wavevector is due to the residual isotropy in the $(r, \phi)$ plane.

\subsection{CMB correlators}

The fact that the primordial power spectra are not solely functions of $k$ means that their relations to the correlators for temperature and polarisation fluctuations of the CMB take more complicated expressions. In this section we state and explain the results, but relegate their explicit derivation to Appendix \ref{app-correlators}. Furthermore we ignore the effect of late-time anisotropy, with which we will deal in section \ref{sec-late}.

\subsubsection{Scalar mode}

For the comoving curvature perturbation this has already been calculated in \cite{Gumrukcuoglu:2007bx}, and we have
\be
C^{(S)XY}_{ll'mm'} = \frac{\delta_{mm'}}{\pi} \int_0^\infty \frac{dk}{k} \Delta^{(S)X}_l (k\Delta\eta) \Delta^{(S)Y}_{l'}(k\Delta\eta) \tilde{P}^{(S)}_{ll'm}(k), \label{CS}
\ee
where $X, Y \in \{T, E\}$, $\Delta^{(S)X}_l$ is the transfer function for temperature or polarisation,\footnote{These are listed in Appendix \ref{app-correlators}, in table \ref{tab-transfer functions}.} $\Delta\eta = \eta_0 - \eta_{LSS}$, and
\be
\tilde{P}^{(S)}_{ll'm}(k) = i^{l-l'} \sqrt{\frac{(2l+1)(2l'+1)(l-m)!(l'-m)!}{(l+m)!(l'+m)!}} \int_{-1}^1 d\mu P^m_l(\mu) P^m_{l'}(\mu) P_\mathcal{R}(k,\mu). \label{PRtilde}
\ee
Note that when $P_\mathcal{R}$ depends only on $k$, due to the orthogonality of the associated Legendre polynomials we have $\tilde{P}^{(S)}_{ll'm}(k) = 2 \delta_{ll'} P_\mathcal{R}(k)$, and the usual isotropic result is recovered.

If $P_\mathcal{R}$ is an even function of $\mu$ (as will be the case here, since parity is still preserved), then only multipoles separated by an even number in $l$-space will be correlated. Also note that the diagonality in $m$ arrises from choosing our isotropic coordinate system to align with the anisotropic one (along with the fact that there is residual isotropy in the $(r, \phi)$ plane), which in general would not be the case and would require a rotation.

\subsubsection{Tensor modes}\label{sec-early, tensor modes}

Unsurprisingly the case for the tensor modes is significantly more complicated due to the fact that the tensor mode functions have spin weight 2.  Here we follow the formalism of \cite{Zaldarriaga:1996xe}, though the derivation in this case is complicated by the fact that we cannot arbitrarily rotate our coordinate system, due to the anisotropy of $P_{+,\times}$. The details again can be found in Appendix \ref{app-correlators}, and the result is
\be
C^{(T) XY}_{ll'mm'} = \frac{\delta_{mm'}}{\pi} \int_0^\infty \frac{dk}{k} \Delta^{(T)X}_l (k\Delta\eta) \Delta^{(T)Y}_{l'}(k\Delta\eta) \tilde{P}^{(T)}_{ll'm}(k), \label{CT}
\ee
where
\begin{align}
\tilde{P}^{(T)}_{ll'm}(k) = i^{l-l'} & \sqrt{\frac{(2l+1)(2l'+1)(l-m)!(l'-m)!}{(l+m)!(l'+m)!}} \nonumber \\
&\times \sum_{i=0}^{l-|m-2|)} \sum_{i'=0}^{l'-|m-2|)} \beta_{ilm}\beta_{i'l'm} \int_{-1}^1 d\mu P^{m-2}_{l-i}(\mu) P^{m-2}_{l'-i'}(\mu) \hat{P}^{(T)}(k,\mu), \label{PTtilde}
\end{align}
and
\be
\hat{P}^{(T)}(k,\mu) = (1 + \sigma_X \sigma_Y (-1)^{l-l'+P} ) (P_+ + P_\times) + (-1)^i (\sigma_X + \sigma_Y (-1)^{l-l'+P}) (P_+ - P_\times), \label{PThat}
\ee
where $\sigma_{T, E} = 1$ and $\sigma_B = -1$, and $P$ is the parity of the primordial power spectrum (even in this case); the constant coefficients $\beta_{ilm}$ are given in Appendix \ref{app-correlators}. 

We have the sum
\be
\sum_{i=0}^{l-|m-2|} \beta_{i,l,m}\beta_{i,l+\Delta l,m} \frac{2(l-i+m-2)!}{(2(l-i)+1)(l-i-m+2)!} = \delta_{0 \Delta l} \frac{2 (l+2)! (l+m)!}{(2l+1)(l-2)!(l-m)!},
\ee
and hence the usual isotropic result is recovered when $\hat{P}^{(T)}$ depends only on $k$.

For modes of the same parity (e.g. $TE$) we again find that only modes separated by an even number in $l$-space are correlated.
However, intriguingly, for modes of opposite parity (e.g. $TB$), for which there is no correlation in the isotropic case, here there is the possibility for correlation of modes separated by an \emph{odd} number in $l$-space. Such a signal is interesting because it is solely due to early-time anisotropy (whereas the other signals discussed also receive contributions from late-time anisotropy).

\subsection{Quasi-de Sitter approximation} \label{sec-qdS approx}

In order to solve the equations of motion for the perturbation mode functions, and thus find the primordial power spectra, we would have to resort to numerical methods in the general case, however they simplify dramatically if we take a quasi-de Sitter approximation in which the energy density is only due to vacuum energy (i.e. the inflaton is frozen). In fact the modes in \eqref{S2v+} decouple, and for the canonically normalised scalar mode we are left with
\be
\omega^2_v = c^2 \left( \frac{q^2}{a^2} + \frac{k_3^2}{b^2} \right) - \frac{\left[ \sqrt{\frac{a^2 b}{c}} \right]''}{\sqrt{\frac{a^2 b}{c} }}. \label{wv}
\ee
The expression for $\omega^2_+$ is significantly more complicated, and so we do not show it here, but instead it can be found in Appendix \ref{app-perturbation theory}. In the isotropic limit these both reduce to $k^2 - \frac{a''}{a}$, as expected for mode functions on de-Sitter space.

\begin{figure}[tp]
   \centering
   \includegraphics[width=0.75\columnwidth]{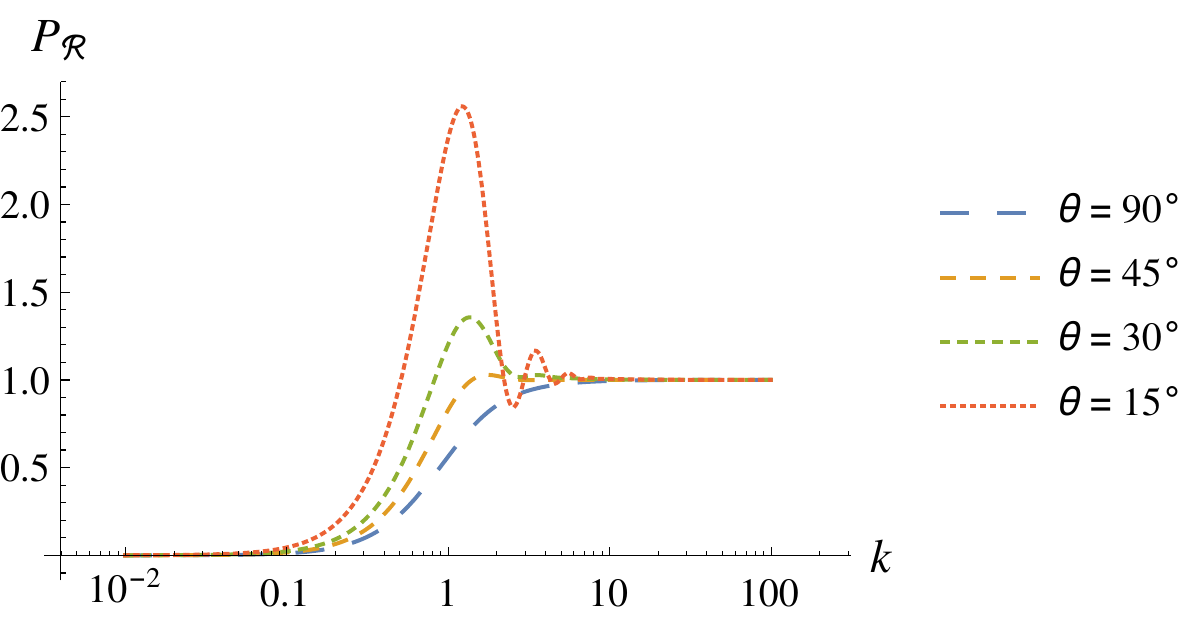} 
   \caption{The primordial power spectrum for the coming curvature perturbation as a function of wavevector ($k = 1$ corresponds to the curvature scale), for various values of the angle $\theta$ between the wavevector and new $z$ dimension. The peak around $k=1$ followed by oscillatory behaviour when $\theta$ is small, is due to the fact that the comoving particle horizon in the $z$ direction is finite (since it starts with finite size), which gives rise to resonance effects.}
   \label{fig-P_R}
\end{figure}

For the rest of this section we will specialise to the case in which the $z$ dimension is the new one (so $\kappa = -1$ and $c = a$) -- performing the analysis for the other cases should be straightforward. In this case solving the equations of motion for the scale factors gives 
\be
a(t) = \sqrt{\frac{3 M_\text{Pl}^2}{V}}\, \mathrm{cosech}(-\eta), \qquad b(t) = \sqrt{\frac{3 M_\text{Pl}^2}{V}} \coth (-\eta), \label{a,b qdS}
\ee
where the conformal time is $\eta = \ln \tanh \left( \frac{1}{2} \sqrt{\frac{V}{3 M_\text{Pl}^2}} t \right)$. Plugging this into \eqref{wv} yields 
\be
\omega_v^2 = k^2 (1-\mu^2 \tanh^2 \eta ) - 2 \, \mathrm{cosech}^2\, \eta -\frac{1}{4} \,\mathrm{sech}^2\, \eta, \label{w2v-approx}
\ee
and through an appropriate set of field redefinitions the equation resulting from this can be recast into the hypergeometric differential equation (see the Appendix \ref{app-solving for modes} for the details).
Thus we can solve it, and evaluate the result at the end of inflation, $\eta \to 0$, to get
\begin{align}
 P_\mathcal{R} \propto& \frac{k^3}{2 \pi^2} \left| \lim_{\eta \to 0}  \frac{v}{a} \right|^2 \\
 &= \frac{k^3}{64 \pi^4} \frac{\sinh^2 \pi \tilde{k} + \cos^2 \pi k\mu}{\sinh \pi \tilde{k}} \left| \Gamma \left(-\frac{1}{4} + \frac{1}{2}k\mu + \frac{1}{2} i \tilde{k} \right) \Gamma \left(-\frac{1}{4} - \frac{1}{2}k\mu + \frac{1}{2} i \tilde{k} \right) \right|^2 \label{P_R exact}\\
 &= \frac{1}{8\pi^2} \frac{\cosh \pi \tilde{k} - \sin \pi k \mu}{\sinh \pi\tilde{k}} \left( 1 + \frac{5}{8} k^{-2} (2\mu^2 -1) + \mathcal{O}\left( k^{-3} \right) \right), \label{P_R large k}
 \end{align}
where $\tilde{k} \equiv k\sqrt{1 - \mu^2} = k \sin \theta$. This is plotted as a function of $k$, for various values of $\mu = \cos \theta$ in figure \ref{fig-P_R}. We see that it oscillates as a function of $k$ when the wavevector has a component along the $z$ direction; mathematically this is due to the $\cos( \pi k\mu)$ term in the numerator of \eqref{P_R exact}, and physically it is due to the fact that the comoving particle horizon in the $z$ direction is finite (since it starts with finite size), which gives rise to resonance effects. The magnitude grows as the wavevector approaches the new $z$ dimension, due to the $\sinh (\pi \tilde{k})$ term in the denominator of \eqref{P_R exact}, and ultimately to the fact that the adiabatic vacuum initial condition blows up there.

The exact effect on the CMB correlators will be discussed in section \ref{sec-results}, but for now let us note that from \eqref{P_R large k} we see that for large $k$ the behaviour can be roughly split into two parts, depending on the value of $\tilde{k} = k \sin \theta$ (the projection of the wavevector along the old dimensions):
\be
P_\mathcal{R} =
\begin{cases}
P_0 (1 + \frac{5}{4} k^{-2} \cos^2\theta ) & \text{for}\quad k \sin \theta > \tilde{k}_t, \\
P_0 \frac{1 - \sin \pi k}{\pi k \sin \theta} & \text{for}\quad k \sin \theta < \tilde{k}_t,
\end{cases} \label{P_R split}
\ee
for some transition value $\tilde{k}_t \approx \frac{1}{4}$, below which $P_\mathcal{R}$ rises sharply. This rise can be attributed to the adiabatic vacuum initial condition we have used, which also blows up in this regime. In reality, immediately after tunnelling the mode function will be in a more complicated state than the adiabatic vacuum, see for example \cite{Hamazaki:1995dy, White:2014aua}, and thus the small $\tilde{k}$ rise in the power spectrum should perhaps be taken with a pinch of salt. In section \ref{sec-results} we shall see that most of the interesting signatures are not in fact too dependent on this.

Now we turn to the $h_\times$ mode. Plugging \eqref{a,b qdS} into \eqref{wx} yields
\be
\omega_\times^2 = \omega_v^2 + \frac{\frac{1}{4} + k^2(1+4\mu^2)}{\left( \frac{1}{4} + k^2(1-\mu^2 \tanh^2 \eta) \right) \cosh^2 \eta} - \frac{3k^2 \mu^2 (\frac{1}{4} + k^2)}{\left[\left( \frac{1}{4} + k^2(1-\mu^2 \tanh^2 \eta) \right) \cosh^2 \eta \right]^2}. \label{w2x-approx}
\ee
The extra terms relative to the scalar mode are only important when $\tilde{k} < 1$, i.e. when the wavevector is pointing significantly in the direction of the new dimension, and so outside this regime we should be safe using the first line of \eqref{P_R split}. The same is true for the $h_+$ mode, though we do not show $\omega_+^2$ here due to its length. Outside of this regime we will need to use the full forms in \eqref{w2x-approx} and \eqref{w2+-approx}, however numerical studies we have done indicate that the behaviour of the tensor power spectra at small $\tilde{k}$ is similar to that of $P_\mathcal{R}$.\footnote{More detailed numerical studies are required to properly investigate the departures of $P_{\times,+}$ from $P_\mathcal{R}$ and each other.} 

The singularity structure of the equation arising from \eqref{w2x-approx} is the same as that of Heun's differential equation and thus it is transformable into that, and in appendix \ref{app-solving for modes} we discuss this and attempt to find the power spectrum $P_\times$. Unfortunately, due to the complexity of Heun functions, we are not able to find a complete, closed form expression, however for small values of $\tilde{k}$ it does show the $\tilde{k}^{-1}$ behaviour exhibited in the scalar mode case, encouraging us to take their behaviour to be the same as \eqref{P_R split} (though with the transition value $\tilde{k}_t$ as a free parameter).

\section{Late-time anisotropy} \label{sec-late}

The background also acquires a non-negligible anisotropy at late times that grows during the periods of radiation and matter domination, and thus we should also include effects due to this when studying the impact on the CMB. In \cite{Graham:2010hh} this was studied extensively for the temperature perturbations, and so here we simply quote their findings. Three effects were found:
\begin{enumerate}[(i)]
\item the surface of last scattering (LSS) is warped, being ellipsoidal in shape rather than spherical;
\item the angle on the LSS from which a given photon originates is offset from the angle at which it is received;
\item the redshift to the LSS is now angle dependent.
\end{enumerate}

The effect of the last two is to shift the coefficients of the multipoles, so that
\be
a_{lm} \to a_{lm} - \Omega_{k0} \sum_{i=-1}^1 \tilde{h}^{l-2i}_i a_{l-2i,m}, \label{late ii iii effects}
\ee
for some coefficients $\tilde{h}^l_i$ which are explained in appendix \ref{app-late}. The first has the effect that the distance to the surface of last scattering acquires an angle dependence:
\be
x = k \Delta\eta \to k \Delta\eta \left( 1 - \frac{8}{45} \sqrt{\frac{\pi}{5}} \Omega_{k0} Y_{20} (\theta) \right),
\ee
and thus the argument of the transfer functions, $k \Delta\eta$, does too. One deals with this (to first order in $\Omega_{k0}$, which is small) by Taylor expanding:
\be
\Delta_l(x) \to \Delta_l(x_0) - \Delta_l'(x_0) x_0 \frac{8}{45} \sqrt{\frac{\pi}{5}} \Omega_{k0} Y_{20} (\theta). \label{late i effect}
\ee
The derivatives of the transfer functions can be found in appendix \ref{app-late}, in table \ref{tab-transfer functions}.

We now need to extend this to deal with polarisation, which is particularly easy, since the same effects apply in that case \emph{apart} from the angle-dependent redshift, and thus the coefficients $\tilde{h}$ in \eqref{late ii iii effects} will be different for polarisation than temperature.\footnote{Note that this effect will still appear in temperature-polarisation cross-correlation, but only through the temperature transfer functions.}

Combing all the effects we have for both scalar and tensor modes, to first order in $\Omega_{k0}$
\begin{align}
C^{XY}_{ll'mm'} =& \frac{\delta_{mm'}}{\pi} \int_0^\infty \frac{dk}{k} \Bigg[ \Delta^X_l \Delta^Y_{l'} \tilde{P}_{ll'm} \nonumber \\
&+ \Omega_{k0} \sum_{i=-1}^{1} \left( \frac{8}{45} \sqrt{\frac{\pi}{5}} f^{l'm}_i \Delta^X_l \Delta'^{Y}_{l'+2i} - \tilde{h}^{l'-2i,m}_i \Delta^X_l \Delta^Y_{l'-2i} + (l \leftrightarrow l') 
\right) \tilde{P}_{l,l'-2i,m} \Bigg], \label{Cllm late combined}
\end{align}
where the coefficients $f$ arise from the $Y_{20}$ spherical harmonic in \eqref{late i effect} and can be found in appendix \ref{app-late}.
This is appropriate both when early time anisotropy effects are not included, and when they are, with the difference being the expression for $\tilde{P}_{ll'm}$ (i.e. \eqref{PRtilde} and \eqref{PTtilde}, or their isotropic simplifications).

\section{Results} \label{sec-results}

Having examined the various ways in which the background anisotropy due to the tunnelling event will affect the CMB correlators, let us now look at the size of the effects and consider their observability. We specialise to the case of a $2+1 \to 3+1$ transition, and use the quasi-de Sitter approximation discussed in section \ref{sec-qdS approx}.

\subsection{Observables}

The key observational difference compared to the standard cosmological model is that certain off-diagonal elements of the correlation matrix are non-zero, and thus we consider the ratio of these to the diagonal elements,
\be
\frac{C_{ll'm}}{C_{llm}}, \qquad \text{ and } \qquad 
\frac{\sum_m C_{ll'm}}{\sum_m C_{llm}} = \frac{C_{ll'}}{C_{ll}}.
\ee
The reason to include the latter is that our results implicitly rely on the coordinate system used to calculate the CMB correlators being aligned with the direction of anisotropy, which in practice is unknown (though of course one could rotate one's coordinate system to find an estimate of the direction of anisotropy).

The leading order effects are all proportional to $\Omega_{k0}$, the value of the curvature parameter today. This can be directly constrained in the same way as isotropic curvature, however there is an even stronger indirect constraint \cite{Graham:2010hh}. The fact that in the $TT$ spectrum modes with $\Delta l = 2$ are correlated means that the CMB temperature quadrupole receives a contribution from the monopole of order $T_0 \Omega_{k0}$; since the temperature fluctuations are $\sim 10^{5}$ times smaller than the monopole, $\Omega_{k0}$ must be correspondingly small in order that we do not observe an anomalously high quadrupole.\footnote{This doesn't, however, give us a precise bound since there can always be cancellations between the different contributions to the quadrupole.}

\subsection{Scalar mode}

Let us first briefly estimate the size of the effect. From \eqref{P_R split} we find
\be
\tilde{P}_{\mathcal{R},ll'm} \propto \delta_{l',l} + k^{-2} \left( c^-_{lm} \delta_{l',l-2} + c^+_{lm} \delta_{l',l+2} + \sum_{i=-\infty}^\infty  \tilde{c}_l \delta_{l',l+2i} \delta_{m,0} \tilde{k}_t (1 - \sin \pi k ) \right) + \mathcal{O}(k^{-3}), \label{PRtilde actual}
\ee
for some coefficients $c$, whose precise values aren't needed in the following paragraph.
To find the CMB correlators we then need to integrate over the transfer functions which all consist of linear combinations of spherical Bessel functions (possibly multiplied by inverse powers of $k\Delta\eta$), and therefore we need to use the integral\footnote{Since the $1-\sin \pi k$ factor oscillates much more rapidly than the Bessel functions, we can just take its average value.}
$\int_0^\infty dk k^{-\nu} j_l (k\Delta\eta)  j_{l'} (k\Delta\eta) \propto (\Delta\eta)^{\nu}$, and so the effects due to early-time anisotropy should be $\sim (\Delta\eta)^2$. Noting that $\Delta\eta \approx 2\sqrt{\Omega_{k0}/z_{LSS}}$, we see that the early-time anisotropy effects should be $\sim 10^{-3} \Omega_{k0}$. This should be compared to the late time anisotropy for which we expect effects $\sim \Omega_{k0}$ when $| l-l'| = 2$.

\begin{figure}[tp]
   \centering
   \includegraphics[width=\textwidth]{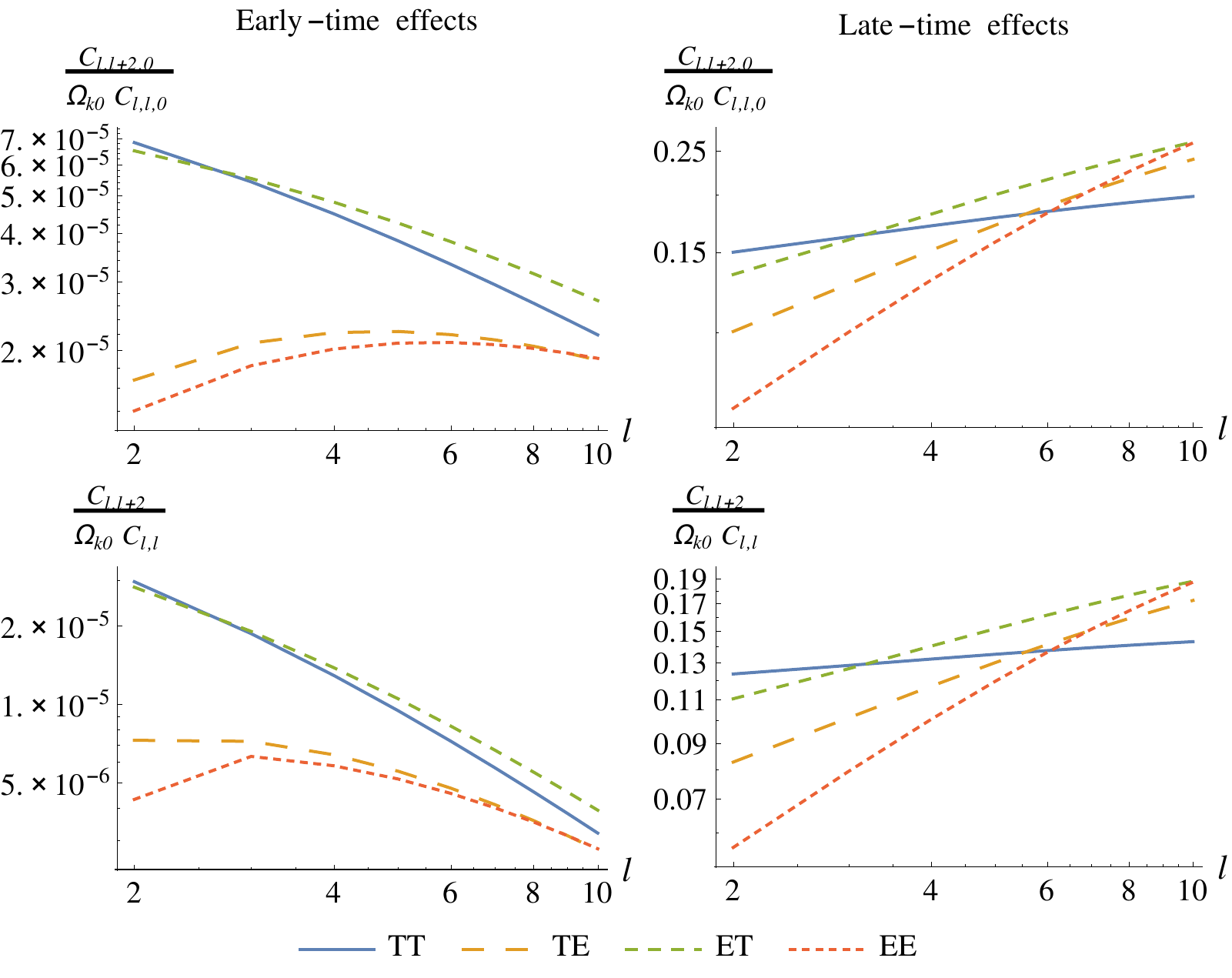} 
   \caption{$\Delta l = 2$ elements of CMB multipole correlation matrix (compared to the $\Delta l = 0$ elements, in units of the anisotropic curvature parameter) due to the scalar mode, comparing the late and early-time effects; top: $m=0$; bottom: sum over all $m$. The early time effects are sub-dominant, and in fact a couple of orders of magnitude smaller than expected}
   \label{fig-scalar results}
\end{figure}

Figure \ref{fig-scalar results} shows the above mentioned observables in units of $\Omega_{k0}$ for each of the two classes of effects (early- and late-time anisotropy). We see that both effects are smaller than one would expect, but that the late-time effect grows with $l$, whilst the early-time effect decreases. Thus it would seem that, as expected, for the scalar mode for $\Delta l = 2$ the late-time effects completely dominate (as predicted in \cite{Graham:2010hh}).

There are, however, observables for which the early-time effects could come dominate: $\Delta l = 2n$ for $n > 1$. The late-time contribution comes from keeping terms higher order in $\Omega_{k0}$ in the expansions in section \ref{sec-late}, and so we would expect them to be $\sim (\Omega_{k0})^n$; including higher order terms in \eqref{P_R split} will generate the desired effect, but at the cost of larger inverse powers of $k$ which results in larger powers of $\Omega_{k0}$ once the $k$ integral is performed, and thus the early-time contribution when the wavevector is not aligned along the new dimension ($\tilde{k} > \tilde{k}_t$) will behave likewise. On the other hand the early-time contribution when the wavevector \emph{is} aligned along the new dimension ($\tilde{k} < \tilde{k}_t$) will generate non-zero correlations for $\Delta l > 2$ which we see from \eqref{PRtilde actual} will \emph{not} come at the expense of higher powers of $\Omega_{k0}$. 

Looking at the actual coefficient in front of these terms once the $k$ integral is performed, and we find that
\be
\frac{C_{l,l+2n,0}}{C_{l,l,0}} \sim 10^{-3} \Omega_{k0} n^{-4}
\ee
(for the $l$ which maximises the effect), which is the desired non-exponential falloff with $n$, and we find that this is similar to or larger than the late-time anisotropy effects for $n \geq 2$. Unfortunately, the effects we are considering are now extremely small ($ \sim 10^{-8}$ for $\Omega_{k0} = 10^{-4}$). As we shall see in the next section, the tensor modes may offer a better opportunity to study the effects of early-time anisotropy.

\subsection{Tensor modes}

\begin{figure}[tp]
   \centering
\includegraphics[width=\textwidth]{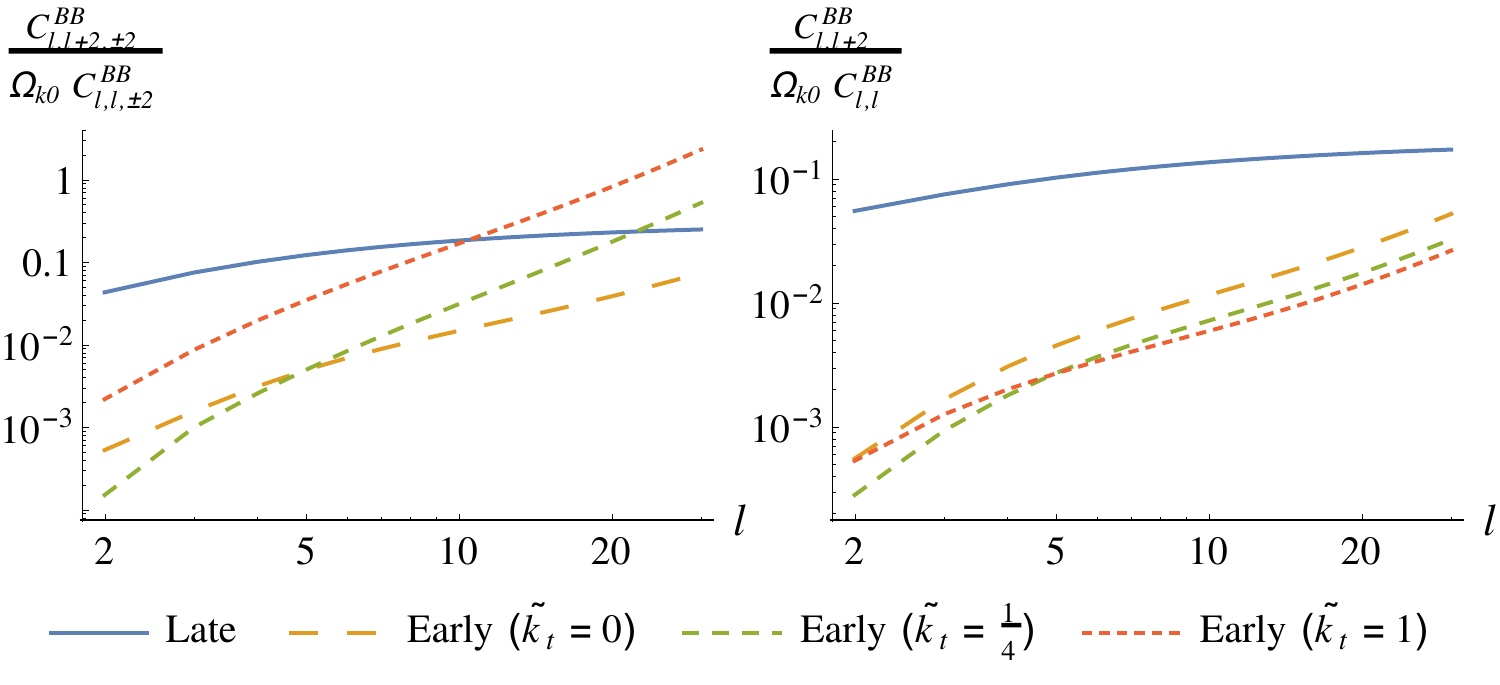}
   \caption{$\Delta l = 2$ elements of $BB$ multipole correlation matrix (compared to the $\Delta l = 0$ elements, in units of the anisotropic curvature parameter), comparing the late and early-time effects; left: $m=\pm 2$; right: sum over all $m$. For the early time effects the tensor power spectra are taken to be \eqref{P_R split}, with $\tilde{k}_t$, the value of $\tilde{k} = k \sin\theta$ at which the power spectrum experiences a sharp rise, a free parameter; $\tilde{k}_t = 0$ corresponds to no sharp rise, $\tilde{k}_t = \frac{1}{4}$ to the same behaviour $P_\mathcal{R}$, and $\tilde{k}_t = 1$ to the earliest possible onset of the rise. The early time effects grow quickly with $l$ becoming several orders of magnitude larger than than expected, and potentially dominant over the late-time effects.}
   \label{fig-tensor results 1}
\end{figure}

\begin{figure}[tp]
   \centering
   \includegraphics[width=\textwidth]{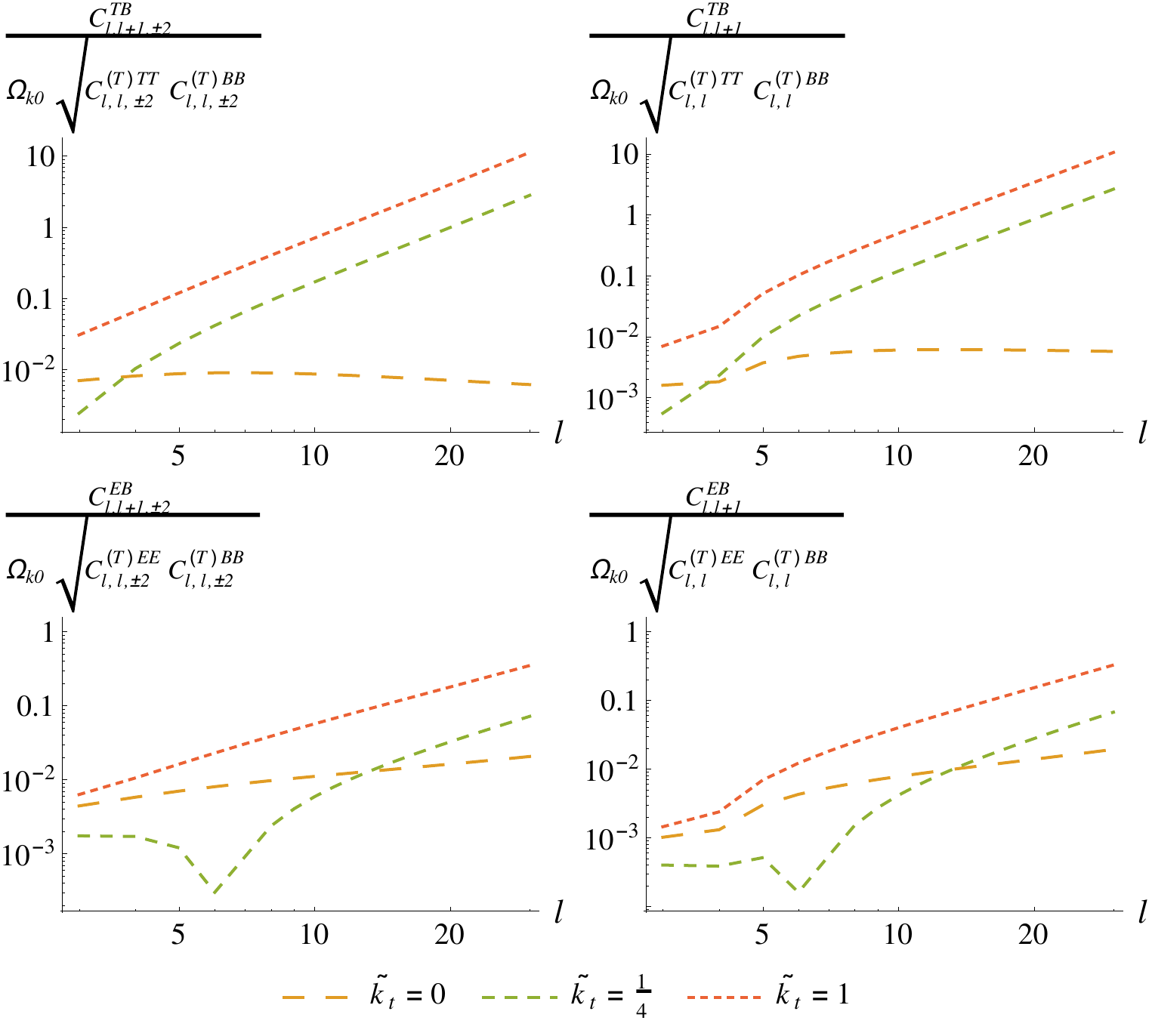} 
   \caption{$\Delta l = 1$ elements of $TB$ and $EB$ multipole correlation matrix (compared to the geometric mean of the $\Delta l = 0$ (tensor) $TT$/$EE$ and $BB$ elements, in units of the anisotropic curvature parameter); left: $m=\pm 2$; right: sum over all $m$. The tensor power spectra are taken to be \eqref{P_R split}, with $\tilde{k}_t$, the value of $\tilde{k} = k \sin\theta$ at which the power spectrum experiences a sharp rise, a free parameter; $\tilde{k}_t = 0$ corresponds to no sharp rise, $\tilde{k}_t = \frac{1}{4}$ to the same behaviour $P_\mathcal{R}$, and $\tilde{k}_t = 1$ to the earliest possible onset of the rise.}
   \label{fig-tensor results 2}
\end{figure}

As explained in section \ref{sec-qdS approx}, for the two tensor modes although we have been unable to derive a complete solution for the primordial power spectra in the quasi-de Sitter approximation, when the wavevector is not aligned along the new dimension ($\tilde{k} > \tilde{k}_t$) their behaviour is the same as the scalar mode, and there are reasons to believe that their behaviour is similar even outside of this regime. Therefore in this section, when examining the early-time anisotropy effects we take the two tensor mode power spectra to be equal, and to have the same form as the scalar mode power spectrum (but with $\tilde{k}_t$ as a free parameter).

Figure \ref{fig-tensor results 1} compares for the $BB$ correlators, the late-time anisotropy effects with the early time anisotropy effect for various values of $\tilde{k}_t$. It shows a stark difference to the scalar mode. We see that, somewhat regardless of the value of $\tilde{k}_t$ the early-time effect grows with $l$, and does so faster than the late-time effect, becoming dominant for $l \sim 10 - 40$,\footnote{We stop the graph at $l = 30$ because above this factors beyond simply the Sachs-Wolfe effect (which is all we have included) come into play when calculating CMB correlators.} even despite the factor of $z_{LSS}^{-1}$ which bedevils the early-time effect. That this behaviour is present even when $\tilde{k}_t = 0$, i.e. when there is no sharp rise in the power spectrum when the wavevector is aligned along the new dimension, indicates that this is not simply due to the initial conditions we have taken for the modes but is a genuine effect (see the discussion towards the end of section \ref{sec-qdS approx}).

This behaviour is also seen in the tensor $TT$ correlators; whilst interesting, this is observationally inaccessible due to the the dominance of the scalar mode. For the $TE$, $ET$, and $EE$ correlators the late-time anisotropy effect grows more quickly (the same is true for the scalar mode: see figure \ref{fig-scalar results}) and so always dominates over the early-time effect.

Now we come to a further observational signature, which was previously mentioned in section \ref{sec-early, tensor modes}: the possibility of $TB$ and $EB$ correlations. Parity remains unbroken however, and so such correlations can only occur between multipoles for which $\Delta l$ is \emph{odd}. To investigate the size of this effect we again compare to a $\Delta l = 0$ correlator in the isotropic case, and as the $TB$ and $EB$ correlators are zero, we choose to compare to $\sqrt{C^{(T)XX}_{ll}C^{(T)BB}_{ll}}$, where $X$ is $T$ or $E$. It should also  be noted that this effect is purely due to early time anisotropy.

The results can be found in figure \ref{fig-tensor results 2}, and we first note that there is dependence on $\tilde{k}_t$, i.e. on the behaviour of the primordial tensor power spectra when the wavevector is aligned along the new dimension; if the power spectrum does not experience a sharp rise ($\tilde{k}_t = 0$) then the $\Delta l = 1$ $EB$ and $TB$ correlators grow quite slowly, with $C^{(T)XB}_{l,l+1}/\sqrt{C^{(T)XX}_{ll}C^{(T)BB}_{ll}}$ reaching $\sim 10^{-2} \Omega_{k0}$ for $l=30$, which is about an order of magnitude smaller than the $C^{BB}_{l,l+2}/C^{BB}_{ll}$ effect. However if $\tilde{k}_t \gtrsim 1/4$ (the value for the scalar mode), then the $EB$ and $TB$ effects can be competitive with the $BB$, $\Delta l = 2$ signal. Unfortunately without having a precise form for the tensor mode primordial power spectra we are unable to determine this effect  precisely and judge how interesting it is.

\section{Conclusions} \label{sec-conc}

In this paper we have examined some of the observational consequences for CMB correlation functions, of a model in which our universe arises out of a tunnelling event from a parent vacuum with a smaller number of large dimensions. Such a scenario is motivated by the string landscape, in which one might expect there to be more vacua with a larger number of compactified (and hence smaller number of large) dimensions.

We have gone beyond previous work \cite{BlancoPillado:2010uw, Adamek:2010sg, Graham:2010hh, Blanco-Pillado:2015dfa} by considering not only the temperature, but also the polarisation power spectra, and also by performing full cosmological perturbation theory at early times in order to investigate the effect not only on the scalar, but also the tensor perturbation modes. 

When analysing the size of the various effects we have specialised to the case of a $2 + 1 \to 3 + 1$ dimension changing pattern; doing the same for the $1+1 \to 3+1$ case should be simple, however it would still be interesting to check if the results are qualitatively the same in that case. In the limit in which the energy density of the universe only comes from vacuum energy and (anisotropic) spatial curvature we have solved exactly for the mode function that becomes the scalar mode at late times; this is also the same as for those which become the tensor modes, whenever the wavevector of the perturbation is not aligned along the `new' dimension. 

For the scalar mode this has allowed us to confirm what was expected, that the effects of early-time anisotropy are sub-dominant to the late-time effects ($\sim 10^{-5} \Omega_{k0}$ versus $\sim 10^{-1} \Omega_{k0}$), for polarisation as well as temperature, and in fact we have shown that the effects are even smaller by a few orders of magnitude than one would naively expect.

Intriguingly for the tensor modes, however, we have found that for the temperature and $B$-mode spectra the early-time effects grow with multipole, such that they can become equal to the late-time effects at around $l \sim 30$; this behaviour is independent of the precise form which we assume for the primordial power spectra when the wavevector is aligned along the `new' dimension. It would be very interesting to extend this analysis to larger $l$ and see to what extent the effects continue to grow, especially their size around $l \sim 100$. We have shown as well that this model produces $TB$ and $EB$ correlations (for multipoles separated by an odd value), although the signal in this case is rather small ($\sim 10^{-2} \Omega_{k0}$).

All of this shows that not only is it testable, a model in which our universe arises out of tunnelling from a vacuum with a smaller number of large dimensions, but also that the effects of the anisotropic primordial power spectrum can also be observed in principle. Unfortunately, given the current and predicted future limits on the tensor-to-scalar ratio \cite{Creminelli:2015oda} it seems unlikely that such a signal would be observable in practice. Given the future prospects for observations of the large-scale distribution of matter in the universe with experiements such as LSST \cite{Ivezic:2008fe}, Euclid \cite{Laureijs:2011gra}, WFIRST \cite{Green:2012mj}, and SKA \cite{Maartens:2015mra}, and especially Hydrogen 21-cm line intensity mapping \cite{Pritchard:2011xb, Bull:2014rha}, it would be very interesting to examine what the effects of the model would be on large-scale structure (though again it seems likely that the effects of the anisotropic primordial power spectrum would be sub-dominant). Indirectly these things thus allow us to probe the landscape hypothesis.
\\
\\
The author is especially grateful to John March-Russell for suggesting the idea behind this work, and for helpful comments on a draft, as well as to Pedro Ferreira for comments.
JHCS is funded by the STFC.

\appendix

\section{Anisotropic perturbation theory} \label{app-perturbation theory}

In this appendix we cover the details of perturbation theory in a universe with background anisotropy both in expansion rate and curvature, but with a residual isotropy in two of the spatial directions. The discussion closely follows \cite{Gumrukcuoglu:2007bx} in which this was done for the case of anisotropic expansion rates, but no spatial curvature.

We assume that inflation is driven by a single, minimally-coupled scalar field, and expand around the background: $g_{\mu\nu} = \bar{g}_{\mu\nu} + h_{\mu\nu}$, $\phi = \bar{\phi} + \varphi$.
Imposing the background field equations and neglecting boundary terms, the action at quadratic order in the perturbations is
\begin{align}
\mathcal{S}^{(2)} = \int d^4x \sqrt{-\bar{g}} \Bigg\{ &\frac{M_\text{Pl}^2}{4} \bigg[  \bar{\nabla}_\lambda h_{\mu\nu} \bar{\nabla}^\nu h^{\mu\lambda} - \bar{\nabla}_\mu h \bar{\nabla}_\nu h^{\mu\nu} + \frac{1}{2} \left( (\bar{\nabla}_\mu h)^2 - (\bar{\nabla}_\lambda h_{\mu\nu})^2 \right) \nonumber \\
&+ \left( (h_{\mu\nu})^2 - \frac{1}{2} h^2 \right) V(\bar{\phi}) \bigg] + h_{\mu\nu} \bar{\nabla}^\mu \varphi \bar{\nabla}^\nu \bar{\phi} - \frac{1}{2} h \left( \bar{\nabla}_\mu \varphi \bar{\nabla}^\mu \bar{\phi} + \varphi V'(\bar{\phi}) \right) \nonumber \\
&- \frac{1}{2} \left( (\bar{\nabla}_\mu \varphi)^2 + \varphi^2 V''(\bar{\phi}) \right) \Bigg\}.
\end{align}

Standard procedure would then be to classify the contributions to $h_{\mu\nu}$ according to their transformation properties under $SO(3)$, but as mentioned in section \ref{sec-early}, due to the symmetries of this problem we instead classify them according to transformation under $SO(2)$. Similarly we leave flexible the definition of conformal time, so that the background metric takes the form $\bar{g}_{\mu\nu} = \mathrm{diag}\left( -c^2, a^2 \gamma_{ij}, b^2 \right)$, where $\gamma_{ij}$ ($i,j = 1,2$) is the metric on the two dimensional space of residual isotropy, whose covariant derivative will be denoted by simply $\nabla_i$. The metric perturbation is decomposed as
\be
h_{\mu\nu} = 
\begin{pmatrix}
-c^2 2\Phi & c^2_{S} \nabla_i B + c^2_{V} B_i & c^2_\chi \partial_3 \chi \\
& a^2 \left( -2 \Psi \gamma_{ij} + 2 \nabla_{(i} \nabla_{j)} E + 2\nabla_{(i} E_{j)} \right) & \tilde{c}^2_{S} \nabla_i \partial_3 \tilde{B} + \tilde{c}^2_{V} \partial_3 \tilde{B}_i \\
& & -b^2 2 \Sigma
\end{pmatrix}, \label{h pert}
\ee
where $(\Phi, \Psi, \Sigma, B, \tilde{B}, E, \chi)$ are scalar perturbations, and $(B_i, \tilde{B}_i, E_i)$ are vector perturbations (with two components each); the vectors are required to be transverse ($\nabla_i B^i = 0$ etc.), which means that they each carry one degree of freedom, and hence we have ten degrees of freedom in total. The derivatives with respect to the third coordinate are chosen for later convenience, and the factors in front of the off-diagonal elements, $c_{S}$ etc., are left free for now (though we shall see later that they drop out of the final expressions).

Let us consider how gauge transformations act on the perturbations. Considering $x^\mu \to x^\mu + \xi^\mu$, and decomposing $\xi^\mu = (\xi^0, \xi^i + \nabla^i \xi, \partial^3 \tilde{\xi})$, we have
\begin{IEEEeqnarray}{rClrClrCl}
\Phi &\to& \Phi  - \frac{c'}{c} \xi^0 - \xi^{0'}, \qquad &\Psi &\to& \Psi + \frac{a'}{a} \xi^0, \qquad &\Sigma &\to& \Sigma + \frac{b'}{b} \xi^0 + \partial_3^2 \tilde{\xi}, \nonumber \\
\chi &\to& \chi + \frac{c^2}{c_\chi^2} \xi^0 - \frac{b^2}{c_\chi^2} \tilde{\xi}', \qquad &B &\to& B + \frac{c^2}{c_S^2} \xi^0 - \frac{a^2}{c_S^2} \xi', \qquad &B_i &\to& B_i - \frac{a^2}{c_V^2} \xi_i, \nonumber \\
\tilde{B} &\to& \tilde{B} - \frac{a^2}{\tilde{c}_S^2} \xi - \frac{b^2}{\tilde{c}_S^2} \tilde{\xi}, \qquad &\tilde{B}_i &\to& \tilde{B}_i - \frac{a^2}{\tilde{c}_V^2}\xi_i, \qquad &E &\to& E - \xi, \nonumber \\
E_i &\to& E_i - \xi_i, \qquad &\varphi &\to& \varphi - \bar{\phi}' \xi^0. &&& \label{gauge transformations}
\end{IEEEeqnarray}
From this we can see which combinations of perturbations can be set to zero, through appropriate gauge transformations. For example, the gauge can be completely fixed by setting $E = E_i = \tilde{B} = \Psi = 0$, as will be done in the following.\footnote{This choice is partially motivated by a desire to avoid the curved two dimensional subspace.}

\subsection{Vector modes}

As usual the modes with different transformation properties will not mix at quadratic order, and we first focus on the vector modes, of which $B_i$ and $\tilde{B}_i$ remain after fixing the gauge. The action (neglecting boundary terms) is
\begin{align}
\mathcal{S}^{(2)}_{\text{vec}} = \frac{M_\text{Pl}^2}{4} \int d\eta d^3x \sqrt{\gamma} \Bigg\{ &\frac{1}{cb} \bigg[ \left( \tilde{c}_V^2 \tilde{B}_{i,3}' - c_V^2 B_{i,3} \right)^2 + 4 \tilde{c}_V^2 c_V^2 \left( \frac{a'}{a} - \frac{\tilde{c}_V'}{\tilde{c}_V} \right) \tilde{B}_{i,3} B_{i,3} \nonumber \\
&+ 2 \tilde{c}_V^4 \Bigg( c^2 V - \left(\frac{\tilde{c}_V'}{\tilde{c}_V} \right)' -2\left(\frac{\tilde{c}_V'}{\tilde{c}_V} \right)^2 +\frac{\tilde{c}_V'}{\tilde{c}_V} \left( \frac{b'}{b} +\frac{c'}{c} \right) - 2 \frac{a'}{a}\frac{b'}{b} \bigg) \left(\tilde{B}_{i,3} \right)^2 \bigg] \nonumber \\
&+ \frac{1}{a^2} \left[ \frac{c}{b} \tilde{c}_V^4 \tilde{B}_{i,3} \left( \Delta - \kappa \right) \tilde{B}_{i,3} - \frac{b}{c} c_V^4 B_i \Delta B_i \right] \Bigg\}, \label{S_vec}
\end{align}
where $\Delta$ is the two dimensional Laplacian on $(r, \phi)$. We expand in eigenfunctions of this (denoting the eigenvalue $-q^2$) and Fourier transform in the $z$ dimension and notice that $B_i$ appears without time derivatives; its equation of motion gives
\be
B_i = \left(\frac{\tilde{c}_V}{c_V}\right)^2 \frac{k_3^2}{\frac{b^2}{a^2}q^2 + k_3^2} \left( \tilde{B}_i' + 2 \left( \frac{\tilde{c}_V'}{\tilde{c}_V} - \frac{a'}{a} \right) \tilde{B}_i \right).
\ee
Substituting this into \eqref{S_vec}, and making the field redefinition
\be
h_\times = \frac{M_\text{Pl}}{\sqrt{2}} \frac{\tilde{c}_V^2}{a^2} \sqrt{q^2 k_3^2} \left[\frac{bc}{a^2} \left( \frac{q^2}{a^2} + \frac{k_3^2}{b^2} \right)\right]^{-\frac{1}{2} } \tilde{B}_i, \label{vector mode transformation}
\ee
then yields \eqref{S2x} after some algebra.

\subsection{Scalar modes}

Of the scalar modes, $\Phi$, $\Sigma$, $B$, $\chi$, and $\varphi$ remain after fixing the gauge; their action is
\begin{align}
\mathcal{S}^{(2)}_\text{sca} = \frac{M_\text{Pl}^2}{4} \int &d\eta d^2q dk_3 \sqrt{\gamma} a^2 b c \Bigg\{ \frac{4}{M_\text{Pl}^2} \bigg[ (\Sigma - \Phi) \varphi V' - \frac{\dot{\bar{\phi}}}{c} \left( (\Sigma + \Phi) \varphi' + \left( c_S^2\frac{q^2}{a^2} B + c_\chi^2\frac{k_3^2}{b^2} \chi \right) \varphi \right) \nonumber \\
&- \frac{1}{2} \left( \frac{1}{c^2} (\varphi')^2 + \left( \frac{q^2}{a^2} + \frac{k_3^2}{b^2} + V'' \right) \varphi^2 \right) \bigg] + 4 \left( \frac{\kappa}{a^2} - V \right)\Phi^2 + \frac{1}{c^2}\frac{q^2}{a^2}\frac{k_3^2}{b^2} \left( c_S^2 B - c_\chi^2 \chi \right)^2 \nonumber \\
&+ \frac{c_S^2}{c^2} \frac{q^2}{a^2} \left( \left( \frac{a'}{a} + \frac{b'}{b} \right) \Phi - \left( \frac{a'}{a} - \frac{b'}{b} \right) \Sigma + \Sigma' \right) B - 2 \frac{\kappa}{a^2} \frac{c_S^4}{c^2}\frac{q^2}{a^2} B^2 \nonumber \\
&- 4 \left( \frac{q^2}{a^2} \Sigma + \frac{2}{c^2} \frac{a'}{a} \Sigma' - 2 \frac{c_\chi^2}{c^2} \frac{k_3^2}{b^2} \frac{a'}{a} \chi \right) \Phi \bigg\}.
\end{align}
$\chi$, $\Phi$, and $B$ only appear as auxiliary fields, and upon integrating them out one reaches \eqref{S2v+}, in which the matrix appearing in the kinetic term is
\be
K = \frac{a^2 b}{c} \left(1 + \frac{2 \kappa \beta}{q^2 \delta^2} \right)^{-1}
\begin{pmatrix}
1 - \frac{2 \kappa}{q^2 \delta^2} \left( \frac{\dot{\bar{\phi}}^2}{2 M_\text{Pl}^2} - \beta \right) & \frac{\dot{\bar{\phi}}}{\delta} - \frac{2 \kappa H_a \dot{\bar{\phi}}}{q^2 \delta^2} \\
& \frac{\dot{\bar{\phi}}^2}{\delta^2} + \frac{2 M_\text{Pl}^2 H_a^2}{\delta^2} - \frac{4 \kappa M_\text{Pl}^2 H_a^2}{q^2 \delta^2}
\end{pmatrix}, \label{K matrix}
\ee
where $\beta = -V +\frac{\kappa}{a^2} - 4 H_a^2 \frac{k_3^2 a^2}{q^2 b^2}$, and $\delta = H_a + H_b + 2 H_a \frac{k_3^2 a^2}{q^2 b^2}$. The second matrix appearing in \eqref{S2v+} is
\be
\tilde{K} = \frac{2 a^2 b}{q^2 \delta^2} \left(1 + \frac{2 \kappa \beta}{q^2 \delta^2} \right)^{-1}
\begin{pmatrix}
-\frac{\dot{\bar{\phi}}}{M_\text{Pl}^2} \bigg\{ \dot{\bar{\phi}} \Big[ q^2 \delta \left(1 + \frac{k_3^2 a^2}{q^2 b^2}\right) &\quad &  -\dot{\bar{\phi}} \bigg\{ q^2 \delta^2 + q^2 (H_a - H_b)\delta \\
- 4 \kappa H_a \frac{k_3^2 a^2}{q^2 b^2} \Big] + \kappa V' \bigg\} & & + \kappa \left[\frac{q^2}{a^2} + 2\beta\right] \bigg\} \\
\\
-2 \dot{\bar{\phi}} k_3^2\frac{a^2}{b^2} \bigg\{ \beta + (3H_a +V') \delta & & q^2 M_\text{Pl}^2 \bigg\{ \frac{q^2}{a^2}\delta - 2 H_a \frac{\kappa}{a^2} \\
- 2 \kappa \frac{H_a}{q^2} (2H_a +V') -2 \dot{\bar{\phi}} (\beta + H_a \delta)\bigg\} & & - 2 (H_a - H_b)(\beta + H_a \delta) \bigg\}
\end{pmatrix}.
\ee
Due to the complexity of the effective mass matrix, $M$, we do not show it here.

The kinetic part of this can be diagonalised (so as to properly identify the mode functions) by performing the field redefinition $h = C H$, where the matrix $C$ satisfies the following conditions
\be
C^\dagger C = K, \qquad \text{ and } \qquad C^\dagger C' - {C'}^\dagger C = \frac{1}{2} \left( \tilde{K} - \tilde{K}^\mathrm{T} \right).
\ee
The action is then
\be
\mathcal{S}^{(2)}_\text{sca} = \frac{1}{2} \int d\eta d^2q dk_3 \sqrt{\gamma} \left\{ {h'}^\dagger h' - h^\dagger \left[ {C^{-1}}^\dagger \left(M + \frac{1}{2} \tilde{K}' \right) C^{-1} -  C'' C^{-1} \right] h
\right\}.
\ee
There are enough conditions to uniquely determine $C$ (up to an overall constant unitary transformation), and in the absence of spatial curvature one can solve this, however in the general case it seems that one may need to resort to numerical methods.

\subsection{Relation to isotropic modes}

In order for the correct interpretation we must relate these modes to their counterparts in the isotropic case ($a = b$, $\kappa = 0$). In longitudinal gauge the metric perturbation takes the following form (for a perturbation with wavevector $(q, 0, k_3)$):
\be
h^\text{iso}_{\mu\nu} = a^2
\begin{pmatrix}
-2 \Phi^\text{iso} & 0 & 0 & 0 \\
& -2 \Psi^\text{iso} + \frac{k_3^2}{k^2} h^\text{iso}_+ & \frac{k_3}{k} h^\text{iso}_\times & - \frac{q k_3}{k^2} h^\text{iso}_+ \\
& &  -2 \Psi^\text{iso} - h^\text{iso}_+ & - \frac{q}{k} h^\text{iso}_\times \\
& & & -2 \Psi^\text{iso} + \frac{q^2}{k^2} h^\text{iso}_+
\end{pmatrix}, \label{h pert iso}
\ee
and the normalised mode functions are
\be
v = a \left( \varphi + \frac{a \bar{\phi}'}{a'} \Psi^\text{iso} \right), \qquad h_+ = \frac{M_\text{Pl}}{\sqrt{2}} a h^\text{iso}_+, \qquad h_\times = \frac{M_\text{Pl}}{\sqrt{2}} a h^\text{iso}_\times.
\ee
From \eqref{gauge transformations} we see that \eqref{h pert iso} can be transformed into \eqref{h pert} (in the isotropic limit) in our gauge ($E = E_i = \tilde{B} = \Psi = 0$) via the following gauge transformation
\be
\xi^0 = - \frac{a}{a'} \left( \Psi^\text{iso} + \frac{1}{2} h^\text{iso}_+ \right), \qquad \xi^2 = -i \frac{k_3}{q k} h^\text{iso}_\times, \qquad \xi = -\frac{q^2 + 2k_3^2}{2q^2 k^2} h^\text{iso}_+, \qquad \tilde{\xi} = \frac{3q^2 + 2k_3^2}{2 q^2 k^2} h^\text{iso}_+.
\ee
(Note that due to the transverse condition on $SO(2)$ vectors ($k_i v^i = 0$), and our choice of wavevector, in this section we have $\xi^1 = 0$ for any transverse $SO(2)$ vector.) Applying this we can relate the dynamical modes to the isotropic modes and hence to the normalised modes:
\be
v = a \left( \varphi + \frac{q^2}{2k^2} \frac{a \bar{\phi}'}{a'} \Sigma \right), \qquad h_+ = -\frac{M_\text{Pl}}{\sqrt{2}} a \frac{q^2}{k^2} \Sigma, \qquad h_\times = -i \frac{M_\text{Pl}}{\sqrt{2}} a \frac{q k_3}{k}\tilde{B}_2.
\ee
Thus we have the correct physical interpretation of our modes in the isotropic limit (which is valid at the end of inflation). In particular notice that the final expression agrees with the isotropic limit of \eqref{vector mode transformation} (up to a phase), and the first two to that which one would get from \eqref{K matrix} in the quasi-de Sitter limit used in the main text.

\subsection{$h_+$ mode function}

For completeness we here present the effective mass term for the $h_+$ mode function in the quasi-de Sitter limit:
\begin{align}
\omega^2_+ = c^2 &\left(1-\frac{2\kappa}{q^2}\right)^{-1} \left[ \left(\frac{\kappa}{2a^2} + \frac{\Delta \mathcal{H}}{\mathcal{H}_a} \frac{c \delta}{\mathcal{H}_a}\right) \frac{q^2}{a^2} - \frac{\Delta \mathcal{H}^2}{\mathcal{H}_a^2} \right] - \frac{\left[\sqrt{\frac{a^2 b \mathcal{H}_a^2}{c^3 \delta^2} \left(1 + \frac{2\beta\kappa}{q^2 \delta^2}\right)^{-1}}\right]''}{\sqrt{\frac{a^2 b \mathcal{H}_a^2}{c^3 \delta^2} \left(1 + \frac{2\beta\kappa}{q^2 \delta^2}\right)^{-1}}} \nonumber \\
&+ \frac{1}{2} \frac{\left[ \frac{a^2 b \mathcal{H}_a^2}{c^3 \delta^2} \left( \frac{c^3 \delta}{\mathcal{H}_a^2} \frac{q^2}{a^2} - 2 c^2 \frac{\kappa}{a^2 \mathcal{H}_a^2} - 2 \Delta \mathcal{H} \left( 1 - \frac{c \delta}{\mathcal{H}_a} \right)  \right) \left(1 + \frac{2\beta\kappa}{q^2 \delta^2}\right)^{-1} \right]'}{\frac{a^2 b \mathcal{H}_a^2}{c^3 \delta^2} \left(1-\frac{2\kappa}{q^2}\right) \left(1 + \frac{2\beta\kappa}{q^2 \delta^2}\right)^{-1}},
\label{w2+-approx}
\end{align}
where $\Delta \mathcal{H} = \mathcal{H}_a - \mathcal{H}_b$ (the calligraphic font represents conformal time Hubble factors), and $\beta$ and $\delta$ are defined below \eqref{K matrix}.

\section{CMB correlators with anisotropic primordial power spectra}\label{app-correlators}

The standard formulae for CMB correlators in terms of primordial power spectra make use of the isotropy of the spectra and so must be generalised for our case. The following two subsections sketch the derivations of \eqref{CS} and \eqref{CT}; the transfer functions and their derivatives can be found in table \ref{tab-transfer functions}.

\begin{table}[htbp]
   \centering
   \begin{tabular}{c c c c}
       & & $\Delta_l(x)$ & $\Delta'_l(x)$ \\
      \hline
	scalar & $T$ & $j_l(x)$ & $\frac{l}{x} j_l(x) - j_{l+1}(x)$ \\
	 & $E$ & $\frac{(l+2)!}{(l-2)!} \frac{1}{x^2} j_l(x)$ & $\frac{(l+2)!}{(l-2)!} \frac{1}{x^2} \left( \frac{l-2}{x} j_l(x) - j_{l+1}(x) \right)$ \\
	 tensor & $T$ & $\frac{1}{2x^2} j_l(x)$ & $\frac{1}{2x^2} \left( \frac{l-2}{x} j_l(x) - j_{l+1}(x) \right)$ \\ 
	  & $E$ & $\left( 1 - \frac{(l+1)(l+2)}{2x^2} \right) j_l(x) + \frac{1}{x} j_{l+1}(x)$ & $(l+1) \left( \frac{1}{x} - \frac{(l+2)(l-2)}{2x^2} \right) j_l(x) - \left( 1 - \frac{l^2 +l -4}{2x^2} \right) j_{l+1}(x)$ \\
	  & $B$ & $\frac{l+2}{x} j_l(x) - j_{l+1}(x)$ & $\left( \frac{(l+2)(l-1)}{x^2} - 1 \right) j_l(x)$
   \end{tabular}
   \caption{Transfer functions and their derivatives; derivatives acting directly on spherical Bessel functions have been removed using the recursion relation $j_l'(x) = \frac{l}{x} j_l(x) - j_{l+1}(x) = j_{l-1}(x) - \frac{l+1}{x} j_l(x)$.}
   \label{tab-transfer functions}
\end{table}

\subsection{Scalar mode}

For the scalar mode this is not so difficult, and in fact the appropriate expression for $TT$ can be found in \cite{Gumrukcuoglu:2007bx}. 
The perturbations projected onto the sky are related to the coming curvature perturbation $\mathcal{R}$ by \cite{Zaldarriaga:1996xe}
\begin{align}
\Delta^{(S)T} &\propto \mathcal{R} \hat{\mathcal{T}}^{(S)} e^{i x \uv{k}\cdot\uv{n}} = 4\pi \sum_{L,M} i^L (2L+1) \hat{\mathcal{T}}^{(S)} j_L(x) Y^*_{LM}(\uv{k}) Y_{LM}(\uv{n}) \mathcal{R}(\mathbf{k})\\
\Delta^{(S)E} &\propto \mathcal{R} \hat{\mathcal{E}}^{(S)} e^{i x \uv{k}\cdot\uv{n}} = 4\pi \sum_{L,M} i^L (2L+1) \hat{\mathcal{E}}^{(S)} j_L(x) Y^*_{LM}(\uv{k}) Y_{LM}(\uv{n}) \mathcal{R}(\mathbf{k}),
\end{align}
where $x = k \Delta \eta$ and the operators are $\hat{\mathcal{T}}^{(S)} = 1$, $\hat{\mathcal{E}}^{(S)} = (1+\partial_x^2)^2 x^2$. In the usual way the CMB perturbations are decomposed into spherical harmonics with coefficients $a_{lm}$, and we have
\begin{align}
C^{(S)XY}_{ll'mm'} &= \langle a^{(S)X}_{lm} a^{(S)Y*}_{l'm'} \rangle = \int \frac{d^3 \mathbf{k}}{(2\pi)^3}  \frac{d^3 \mathbf{k}'}{(2\pi)^3} d\Omega_{\uv{n}} d\Omega_{\uv{n}'} Y^*_{lm}(\uv{n}) Y_{l'm'}(\uv{n}') \langle \Delta^{(S)X}(\mathbf{k}, \uv{n}) {\Delta^{(S)Y}(\mathbf{k}', \uv{n}')}^* \rangle \\
&\propto \sum_{L,M,L',M'} (4\pi)^2 i^{L-L'} (2L+1)(2L'+1) \int \frac{dk}{k} \Delta^{(S)X}_L(x) \Delta^{(S)Y}_{L'}(x) \nonumber \\ 
&\times \int \frac{d\Omega_{\uv{k}}}{(2\pi)^3} P_{\mathcal{R}}(\mathbf{k}) Y^*_{LM}(\uv{k}) Y_{L'M'}(\uv{k}) \left[ \int d\Omega_{\uv{n}} Y^*_{lm}(\uv{n}) Y_{LM}(\uv{n}) \right] \left[ \int d\Omega_{\uv{n}'} Y_{l'm'}(\uv{n}') Y^*_{L'M'}(\uv{n}') \right],
\end{align}
where $\Delta^{(S)X}_l = \hat{\mathcal{X}}^{(S)} j_l$, and we have used the definition of the power spectrum in \eqref{power spec def}. The angular integrals over $\uv{n}, \uv{n}'$ can be trivially performed using the orthogonality of the spherical harmonics; due to residual isotropy in the $(r, \phi)$ plane, the power spectrum is independent of the azimuthal angle of $\mathbf{k}$ and thus the integral over this variable can easily be performed by rewriting the spherical harmonics using $Y_{lm}(\theta, \phi) = \sqrt{\frac{(2l+1)(l-m)!}{4\pi (l+m)!}} P^m_l(\mu=\cos \theta) e^{i m \phi}$, yielding a term proportional to $\delta_{mm'}$. Thus, when all is said and done, one arrives at \eqref{CS}.

\subsection{Tensor modes}

The tensor modes are necessarily more complicated as the tensor perturbations have spin-weight 2 and hence are not rotationally invariant. We follow the formalism of \cite{Zaldarriaga:1996xe}; the perturbations projected onto the sky are
\begin{align}
\Delta^{(T)T} &\propto \frac{1}{\sqrt{2}} (1- \tilde{\mu}^2) \left( (\psi_+ - i \psi_\times) e^{2i \tilde{\phi}} + (\psi_+ + i \psi_\times) e^{-2i \tilde{\phi}} \right) \hat{\mathcal{T}}^{(T)} e^{i x \uv{k}\cdot\uv{n}} \\
\Delta^{(T)E} &\propto \frac{1}{\sqrt{2}} (1- \tilde{\mu}^2) \left( (\psi_+ - i \psi_\times) e^{2i \tilde{\phi}} + (\psi_+ + i \psi_\times) e^{-2i \tilde{\phi}} \right) \hat{\mathcal{E}}^{(T)} e^{i x \uv{k}\cdot\uv{n}} \\
\Delta^{(T)B} &\propto \frac{1}{\sqrt{2}} (1- \tilde{\mu}^2) \left( (\psi_+ - i \psi_\times) e^{2i \tilde{\phi}} - (\psi_+ + i \psi_\times) e^{-2i \tilde{\phi}} \right) \hat{\mathcal{B}}^{(T)} e^{i x \uv{k}\cdot\uv{n}},
\end{align}
where $x = k \Delta \eta$, the operators are $\hat{\mathcal{T}}^{(T)} = 1$, $\hat{\mathcal{E}}^{(T)} = (-12 + x^2(1 - \partial_x^2) - 8x \partial_x)$, $\hat{\mathcal{B}}^{(T)} = (8x + 2x^2 \partial_x)$, and $(\tilde{\mu} = \cos \tilde{\theta} = \uv{k}\cdot\uv{n}, \tilde{\phi})$ are the coordinates of the direction of observation, $\uv{n}$, in a coordinate system in which $\uv{k} = \mathbf{z}$. Then in the usual way we have
\begin{align}
C^{(T)XY}_{ll'mm'} &= \langle a^{(T)X}_{lm} a^{(T)Y*}_{l'm'} \rangle = \int \frac{d^3 \mathbf{k}}{(2\pi)^3} \frac{d^3 \mathbf{k}'}{(2\pi)^3} d\Omega_{\uv{n}} d\Omega_{\uv{n}'} Y^*_{lm}(\uv{n}) Y_{l'm'}(\uv{n}') \langle \Delta^{(T)X}(\mathbf{k}, \uv{n}) {\Delta^{(T)Y}(\mathbf{k}', \uv{n}')}^* \rangle \\
&\propto \frac{1}{2} \int \frac{dk}{k} \frac{d\Omega_{\uv{k}}}{(2\pi)^3} \Big\{ (P_+ + P_\times) \left[ \left( \hat{\mathcal{X}} I^\Omega_{lm+} \right) \left( \hat{\mathcal{Y}} \left(I^{\Omega}_{l'm'+}\right)^* \right) +\sigma_X \sigma_Y \left( \hat{\mathcal{X}} I^\Omega_{lm-} \right) \left( \hat{\mathcal{Y}} \left(I^{\Omega}_{l'm'-}\right)^* \right) \right] \nonumber \\
& \qquad\qquad + (P_+ - P_\times) \left[ \sigma_X \left( \hat{\mathcal{X}} I^\Omega_{lm-} \right) \left( \hat{\mathcal{Y}} \left(I^{\Omega}_{l'm'+}\right)^* \right) + \sigma_Y \left( \hat{\mathcal{X}} I^\Omega_{lm+} \right) \left( \hat{\mathcal{Y}} \left(I^{\Omega}_{l'm'-}\right)^* \right) \right] \Big\}
\end{align} \label{CT app}
where $\sigma_{T,E} = 1$ and $\sigma_B = -1$, and
\be
I^\Omega_{lm\pm} = \int d\Omega_{\uv{n}} Y^*_{lm}(\uv{n}) (1-\tilde{\mu}^2) e^{ix \uv{k}\cdot\uv{n}} e^{\pm 2i \tilde{\phi}}.
\ee
Note that $I^\Omega_{l,m,-} = (-1)^{l+m} \left(I^{\Omega}_{l,-m,+}\right)^*$, and so in fact we will only work with $I^\Omega_{lm+}$ and drop the $\pm$ subscript, and we also take $m \geq 0$ since $I^\Omega_{l,-m} = \left( I^\Omega_{l,m}(- \mathbf{k}) \right)^*$.

In the isotropic case these angular integrals are simple to perform as we are free to rotate $\mathbf{k}$ such that it is along the $\mathbf{z}$ direction, and the coordinates of $\uv{n}$ are simply $(\tilde{\mu}, \tilde{\phi})$. However in the absence of isotropy of $P_{+,\times}$ we cannot do this and must keep the directions of $\mathbf{k}$ and $\uv{n}$ general. Decomposing the plane wave, we have
\be
I^\Omega_{lm} = 4\pi \sum_{L,M} i^L (2L+1) j_L(x) Y^*_{LM}(\uv{k}) \int d\Omega_{\uv{n}} Y^*_{lm}(\uv{n}) Y_{LM}(\uv{n}) (1-\tilde{\mu}^2) e^{2i \tilde{\phi}}, \label{IOmega 1}
\ee
and a little geometry reveals
\begin{align}
e^{2i \tilde{\phi}} = (1 - \tilde{\mu}^2)^{-1} \bigg( &\frac{1}{2} (-1 + 3 c_n^2) s_k^2 - 2 s_n s_k c_n \left[c_k \cos(\phi_n - \phi_k) + i \sin (\phi_n - \phi_k) \right] \nonumber \\
&+ s_n^2 \left[ \left( 1 - \frac{1}{2} s_k^2 \right) \cos \left( 2(\phi_n - \phi_k)\right) + i c_k \sin \left( 2(\phi_n - \phi_k)\right) \right] \bigg),
\end{align}
where $c_n = \cos \theta_n$, $s_n = \sin \theta_n$, with $(\theta_n, \phi_n)$ the coordinates of $\uv{n}$, and similarly for $\mathbf{k}$. Plugging this into \eqref{IOmega 1} and rewriting the spherical harmonics in terms of associated Legendre polynomials, we have
\be
I^\Omega_{lm} = \frac{1}{2} \sqrt{\frac{\pi (2l+1)(l-m)!}{(l+m)!}} \sum_L (2L+1) i^L j_L(x) \sum_{\rho = -2}^2 \frac{(L-(m+\rho))!}{(L+(m+\rho))!} I_{\rho,l,L,m} f_\rho (\mu_k) P^{m+\rho}_L (\mu_k) e^{-im \phi_k},
\ee
where
\begin{IEEEeqnarray}{rClrCl}
I_{0,l,L,m} &=& \int_{-1}^1 d\mu\, \mu^2 P^m_l(\mu) P^m_L(\mu), & f_0 &=& 1 - \mu_k^2 \\
I_{\pm1,l,L,m} &=& \int_{-1}^1 d\mu\, \mu \sqrt{1- \mu^2} P^m_l(\mu) P^{m \pm1}_L(\mu), \qquad & f_{\pm1} &=& -2 \sqrt{1-\mu_k^2} (\mu_k \mp 1) \\
I_{\pm2,l,L,m} &=& \int_{-1}^1 d\mu\, (1-\mu^2) P^m_l(\mu) P^{m\pm2}_L(\mu), & f_{\pm2} &=& \frac{1}{2} (1 \mp \mu_k)^2.
\end{IEEEeqnarray}
Using recurrence relations for the associated Legendre polynomials, and their orthogonality properties, the integrals can be evaluated exactly. Doing this and examining the form of $I^\Omega_{lm}$ for various values of $l$ and $m$ reveals that it takes a surprisingly simple form:
\be
I^\Omega_{l,m} = -4 i^l \sqrt{\frac{\pi (2l+1)(l-m)!}{(l+m)!}} \tilde{\Delta}_l(x) e^{-i m \phi_k} \sum_{i=0}^{l-|m-2|} \beta_{ilm} P^{m-2}_{l-i} (\mu_k), \label{IOmega 2}
\ee
where the coefficients $\beta$ are given by
\begin{align}
\beta_{0lm} &= l(l-1)(l-m+1)(l-m+2), \\
\beta_{i>0,lm} &= (-1)^{i+1} 2m \left(\frac{1}{2}(m+1)i(i-(2l+1)) + (m-1) + l(l+1) \right),
\end{align}
and
\be
\tilde{\Delta}_l (x) = \sum_{i=-1}^1 \frac{2l+4i+1}{2^{|i|} (2(l+i)-1)(2(l+i)+1)(2(l+i)+3)} j_{l+2i}(x) = \frac{1}{2x^2} j_l(x).
\ee
Finally, plugging \eqref{IOmega 2} into \eqref{CT app} and writing $\hat{\mathcal{X}} \left(\frac{1}{2x^2} j_l(x)\right) = \Delta^X_l(x)$, one recovers \eqref{CT}.

\section{Solving for the mode functions in the quasi-de Sitter approximation}\label{app-solving for modes}

\subsection{Scalar mode}

In order to find the behaviour of the scalar mode function, and hence the power spectrum of the comoving curvature perturbation, we need to solve the following equation
\be
v'' + \left( k^2 (1-\mu^2 \tanh^2 \eta ) - 2 \, \mathrm{cosech}^2\, \eta -\frac{1}{4} \,\mathrm{sech}^2\, \eta \right) v = 0, \label{scalar mode equation}
\ee
where $\eta \to -\infty$ corresponds to the tunnelling event, and $\eta \to 0$ to the end of inflation. Changing variables to $T = \tanh^2 \eta$ and $\Phi = \sqrt{T} (1-T)^{\frac{1}{2}i \tilde{k}} v$, where $\tilde{k} = k \sqrt{1-\mu^2}$, turns this into
\be
T(1-T) \Phi'' + \left(-\frac{1}{2} + \left(-\frac{1}{2} + i \tilde{k} \right) T \right) \Phi' - \left( -\frac{1}{4} k^2 + \frac{1}{4} i \tilde{k} + \frac{1}{16} \right) \Phi = 0,
\ee
which we recognise as the hypergeometric equation, and thus the general solution to \eqref{scalar mode equation} is 
\begin{align}
v(\eta) = (\mathrm{sech}^2 \eta)^{-\frac{1}{2} i \tilde{k}} \Bigg[ &\frac{c_1}{\tanh \eta}\, {}_2F_1 \left(-\frac{1}{4} + \frac{1}{2} \tilde{q}, -\frac{1}{4} - \frac{1}{2} \tilde{q}^*; \frac{1}{2}; \tanh^2 \eta \right) \nonumber \\
& + c_2 \tanh \eta\, {}_2F_1 \left(\frac{5}{4} + \frac{1}{2} \tilde{q}, \frac{5}{4} - \frac{1}{2} \tilde{q}^*; \frac{5}{2}; \tanh^2 \eta \right) \Bigg],
\end{align}
where $\tilde{q} = k \mu + i \tilde{k}$. Setting this equal to the adiabatic vacuum at early times then allows us to fix the constants and we find
\be
c_1 = \frac{e^{i \tilde{k} \ln 2}}{\sqrt{2 \tilde{k}}} \frac{\Gamma\left(-\frac{1}{4} + \frac{1}{2} \tilde{q}\right) \Gamma\left(-\frac{1}{4} - \frac{1}{2} \tilde{q}^*\right)}{\Gamma\left(-\frac{1}{2}\right) \Gamma\left(i \tilde{k}\right)} \frac{\cos \left(\pi k \mu \right) + i \sinh \left( \pi \tilde{k} \right)}{2i \sinh \left(\pi \tilde{k}\right)}.
\ee
The value of $c_2$ is unimportant as we have
\be
 P_\mathcal{R} \propto \frac{k^3}{2 \pi^2} \left| \lim_{\eta \to 0}  \frac{v}{a} \right|^2 = \frac{k^3}{2 \pi^2} \left| \lim_{\eta \to 0}  (v \sinh\eta) \right|^2 = \frac{k^3}{2 \pi^2} \left| c_1 \right|^2\\
\ee
since ${}_2F_1(a,b;c;0) = 1$. This gives \eqref{P_R exact}, and to get \eqref{P_R large k} the gamma functions must be expanded using
\be
\frac{\Gamma(z + a)}{\Gamma(z + b)} \sim z^{a-b} \sum_{i=0}^\infty \frac{G_i(a,b)}{z^i},
\ee
where $G_0 = 1$, $G_1 = \frac{1}{2}(a-b)(a+b-1)$, and $G_2 = \frac{1}{12} \binom{a-b}{2} \left( 3(a+b-1)^2 - (a-b+1) \right)$.

\subsection{Tensor modes}

As mentioned in section \ref{sec-qdS approx}, when $\tilde{k} \gtrsim 1$ the equations for $h_{+,\times}$ take the same form as that for $v$. The full forms for \eqref{w2x-approx} and \eqref{w2+-approx} are complicated, yet it turns out that the equation for $h_\times$ can be transformed into Heun's differential equation. This can be seen by first performing the transformations of the previous subsection, and seeing that the resulting equation has regular singularities at $T \in \{0, 1, a, \infty \}$, where $a = \frac{1}{\mu^2} \left(1 + \frac{1}{4k^2} \right)$, which is precisely the singularity structure of Heun's equation. Performing the further transformation $h \to (a - T)^c h$ with $c = \frac{1}{2}\left(1 + i \sqrt{\frac{4-a}{a-1}} \right)$, then explicitly transforms it into this.

The general solution would then be a sum of two Heun functions with appropriate parameters; to apply the boundary conditions at $\eta = -\infty \implies T = 1$, we need to evaluate the Heun function at argument unity, which requires a connection formula relating solutions regular around $T = 0$ to those regular around $T = 1$. Unfortunately no such formula is known for the general case and so we cannot provide a full solution for the power spectrum; however some results are known \cite{Schafke_Schmidt} which allows us to say that the behaviour for $\tilde{k} \ll 1$ is the same as for the scalar mode, i.e. $P_\times \propto \tilde{k}^{-1}$.

\section{Coefficients and functions appearing in the late-time anisotropy effects}\label{app-late}

Here we state the coefficients appearing in \eqref{Cllm late combined}, the expression for the effects of late-time anisotropy. For more details see \cite{Graham:2010hh}.

The warping of the last-scattering surface leads to the (previously isotropic) transfer functions taking on terms which are proportional to $Y_{20}$, and since each transfer function is multiplied by $Y_{lm}$ in the formula for $C_{ll'mm'}$ these lead to the following coefficients:
\be
Y_{20} Y_{lm} = \sum_{i=-1}^1 f^{lm}_i Y_{l+2i,m},
\ee
where
\begin{align}
f^{lm}_{-1} &= \sqrt{\frac{5}{\pi}} \frac{3}{4}\frac{1}{2l-1} \sqrt{\frac{(l - m - 1) (l - m) (l + m - 1) (l + m)}{(2l-3) (2l+1)}} \\
f^{lm}_0 &= \sqrt{\frac{5}{\pi}} \frac{l^2 + l - 3 m^2}{2 (2 l - 1) (2 l + 3)} \\
f^{lm}_1 &= \sqrt{\frac{5}{\pi}} \frac{3}{4}\frac{1}{2l+3} \sqrt{\frac{(l - m + 1) (l - m+2) (l + m + 1) (l + m+2)}{(2l+1) (2l+5)}}.
\end{align}

The offset of the angle on the LSS at which a given photon is received generates terms which look like
\be
\sin(2 \theta) \partial_\theta Y_{lm} (\theta, \phi) = \sum_{i=-1}^1 g^{lm}_i Y_{l+2i,m},
\ee
where the coefficients are
\be
g^{lm}_{-1} = -\frac{8}{3} (l+1) f^{lm}_{-1}, \qquad g^{lm}_0 = -4 f^{lm}_0, \qquad g^{lm}_1 = \frac{8}{3} l f^{lm}_1.
\ee

If the angle dependence of the redshift is not included, i.e. we are dealing with polarisation, then the coefficients $\tilde{h}^{lm}_i$ appearing in \eqref{late ii iii effects} are simply given by $\frac{1}{15} g^{lm}_i$, whereas if the angle dependent redshift is also included then they are given by $\frac{8}{15}\sqrt{\frac{\pi}{5}} f^{lm}_i + \frac{1}{15} g^{lm}_i$.

\bibliographystyle{JHEP}
\bibliography{aniso}

\end{document}